\documentclass[12pt]{article}
\usepackage{graphicx}
\usepackage{amsmath}

\begin{document}

\title{Double scouring by turbulent jets downstream of a submerged sluice gate}

\author{ I. Bove$^a$, D. Acosta$^a$, N. Guti\'{e}rrez$^a$, \\  V.
Guti\'{e}rrez$^a$,  L. G. Saras\'{u}a$^b$  \\
{\small $^a$Instituto de F\'{\i}sica, Facultad de Ingenier\'{\i}a,
Universidad de la
Rep\'{u}blica,  Uruguay}\\
{\small $^b$Instituto de F\'{\i}sica, Facultad de Ciencias, Universidad de la
Rep\'{u}blica, Uruguay}}

\maketitle

\baselineskip=20pt
\renewcommand{\baselinestretch}{1.2}

\begin{abstract}
In this work we study the scour produced by a jet downstream of a
submerged sluice gate on a   non-cohesive particles sediment bed. The
experiments were performed for various values of sill heights and
fluid depths. New regimes were observed in which two holes are
simultaneously developed. We identified the origins of the two holes and
showed that they are produced by different scouring mechanisms. The
dependence of the position of the holes with the approach fluid depth
and the gate opening were studied and expressed in terms of adequate
non-dimensional numbers.
\end{abstract}


\section{Introduction}

The interaction between a moving fluid and a sediment bed is a
problem of great interest due to the physical processes that are
involved, as well as for its relevance in engineering applications.
The scouring that takes place in the
vicinity of hydraulic structures is particularly interesting, owing
that the erosion can
compromise the safety of such devices. The behavior of the
scouring produced by a jet that emerges downstream of a sluice
gate has been studied in a series of works(  Rajaratnam, 1981 , Chaterjee {\it et al.},1994, Hogg {\it et al.},1997, Bey {\it et al.},1997, Balachandar {\it et al.},2000).

These studies showed that at an initial stage the scour develops
rapidly but as time goes the erosion  enters in a stage in
which the bedform changes very slowly. The details of this evolution
depend on flow field, the geometry, and the characteristics of the
grains. In some works it has been suggested that,
 at a given geometry, the profiles of the scour hole at different
times are similar (Rajaratnam,1981). This topic has been controversial since
the results
 obtained in other works do not support this behavior. Most of these
cited works focused their investigations to study empirically the
relation between the
  bedform, the flow and the properties of the grains. Efforts to
predict the profile of the eroded bed from first principles have been
limited, with the
  exception of the works of Hogg {\it et al.}(1997) and Hill and
Younkin (2009). In spite of the
advances that have been accomplished about the relation between the erosion
and the transport
   of sediments, the basic mechanism of scouring is still not fully
understood, due to the complex nature of the flow fields  (Hogg {\it et al.},1997, Bey {\it et al.},1997).
It is thus of interest the investigation of this subject considering the
different ways in which the erosion can take place. In the present
work
 we study the evolution of a bed of non-cohesive particles downstream
of a submerged sluice gate.
We obtained different regimes in which  two scouring holes
produced by distinct mechanisms are observed. We investigated the origin
of these bedform changes and the interactions between the bed and the flow.
The work is organized as follows: In section 2 we describe the
experimental setup. In section 3 the experimental results are
 presented, describing the holes characteristics and their dependence
with the parameters of the flow. In section 4 we study
 the structure of the flow and the origin of the scour. In section 5
the results are discussed and in section 6 we give a summary of the
work.

\section{Experimental setup}

The experimental work was performed in a recirculating open channel of
3 m long, 0.19 m wide, and 0.3 m of maximum height. With the use of a
pump, the fluid was withdrawn from a discharge tank and conducted to a
provision tank, from where the fluid enters in the channel. The caudal
discharged $Q$ was measured with a flowmeter, and was maintained
constant (with the help of a globe valve) at the value $Q$ = 1 l/s in
all the experiments. In order to reduce the turbulence level, the
discharge in the provision tank was performed at its bottom. In
addition, a flow disperser and grids were implemented to reduce the
kinematical energy of the flow.
A gate located at the end of the channel was used to control the water
depth. The fluid depth $H$ upstream the gate was taken between 3.0 and
5.0 cm. The inclination of the channel is variable. It is adjusted in
order to maintain constant the water depth along the channel.
At the entrance of the channel  a device to destroy larger
vortices was placed. Also, just before the gate at the end of the channel, a gap
is arranged on the floor of the channel to reduce the impact of the
gate upstream. In the studied area, in the center of the channel,
variation of velocity with height was analyzed for a rigid flat bed
without sediment. It was observed that the flow was developed and it
obeys the wall law. The sketch of the experimental arrangement is
given in Fig. 1.
A cohesive particles bed, 2.2 cm deep, was placed at the
whole length of the channel floor. The particles were glass spheres
with diameter between 420 and 840 $\mu$m, with an averaged diameter
$d_p$ = 630 $\mu$m.
A sluice gate of variable opening was installed in the middle of the
channel. The opening of the gate $h$ was varied between 19 mm and
24 mm over the particle bed.
The flow field was visualized using different procedures as PIV
(Particle Image Velocimetry), and ink injection. In the case of PIV,
polyamide particles with 50 $\mu$m diameter were used. A laser (500
mW) and  spherical lens were used to create a light sheet of 1 mm
thick. To optimize the PIV, the glass spheres were stained in black.
This avoided the reflection of the laser light from the spheres, which
improves the quality of the images and PIV measurements.

\begin{figure}
\begin{center}
\includegraphics[width=1\textwidth]{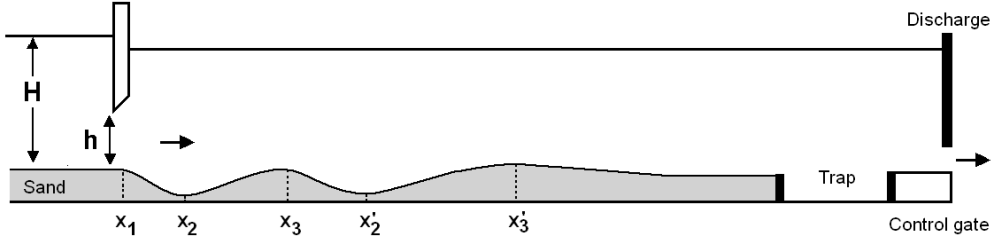}
\end{center}
\caption{Schematic of experimental setup and localization of
characteristics points of the holes.}
\label{setup}
\end{figure}

\section{Experiments}
\subsection{Scour profiles}
In our experiments, the sand bed extends at both sides of the gate. As
a consequence, the scour can begin upstream of the gate. Contrary to
what is usually observed, two holes
of different origin were found to occur. In the schematic of Fig. 1 we
define parameters that determine the position of the first and
second holes. Here, the $x_i$ represent the following positions:
$x_1$ beginning of the first hole, $x_2$  position at which the first
hole achieved the maximum depth, $x_3$ end of the first hole, $x'_1$
beginning of the second hole, $x'_2$ maximum depth of the second hole,
$x'_3$ end of the second hole. We note that the end of the first hole
coincides with the beginning of the second, i.e. $x_3=x'_1$ . All
these
 positions are measured from the localization of the sluice gate.
We also define $S$ and $S'$, which represent the depths of the first
and second holes respectively.
In the present experiments, the profiles did not reach an steady
regime. We began the experiments with a flat bed sand and they
were carried out during a period of 10 minutes. After this, the
position and depths of the holes were measured. These data are those
shown in Figs. 2-5.

\begin{figure}
\begin{center}
\includegraphics[width=1\textwidth]{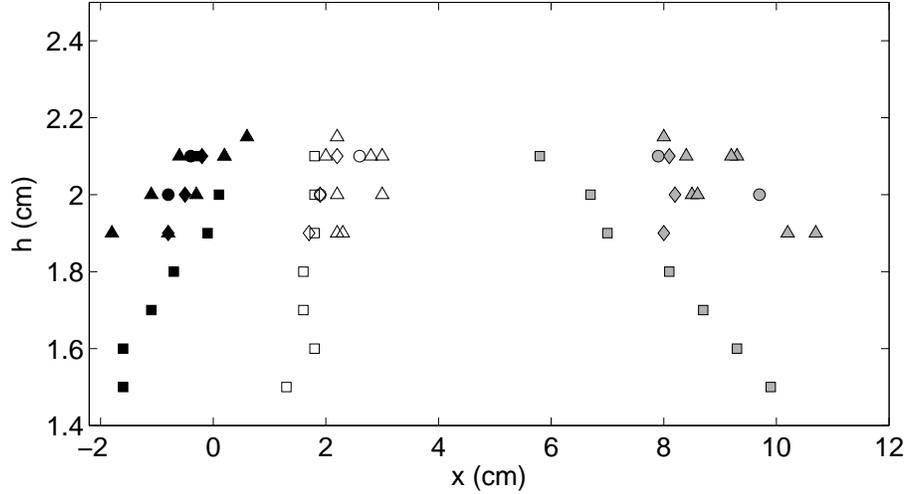}
\end{center}
\caption{Values for the positions of beginning (black symbol), maximum
depth (white) and end (gray) of the first hole, for
 different values of the opening $h$ and the fluid depth $H$, with:
square $H$ = 3 cm, diamond $H$ = 3.5 cm, triangle $H$ = 4 cm and
circle $H$ = 4.5 cm. }
\end{figure}

\begin{figure}
\begin{center}
\includegraphics[width=1\textwidth]{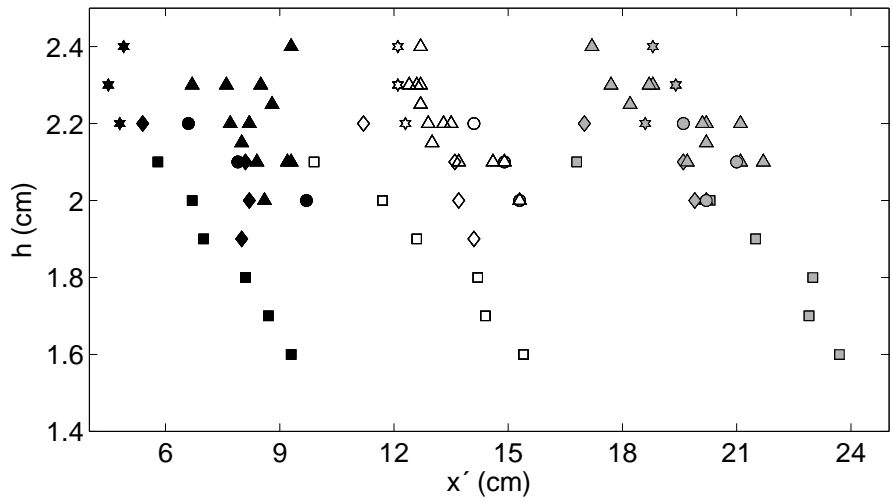}
\end{center}
\caption{Same as figure 2 for the second hole, with: square $H$ = 3
cm, diamond $H$ = 3.5 cm, triangle $H$ = 4 cm, circle $H$ = 4.5 cm and
hexagon $H$=5 cm. }
\end{figure}

\begin{figure}
\begin{center}
\includegraphics[width=1\textwidth]{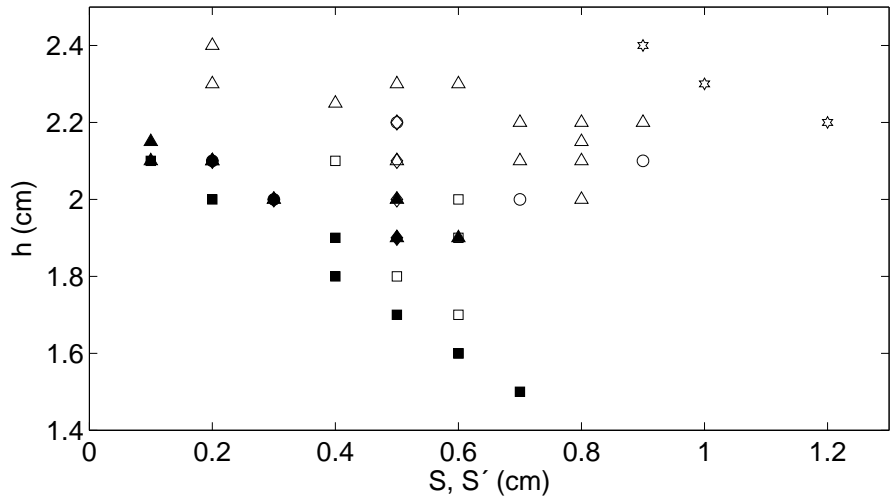}
\end{center}
\caption{Depths $S$,$S'$ of the first hole (black symbols) and second
hole (white symbols) for different gate openings $h$, with: square $H$
= 3 cm, diamond $H$ = 3.5 cm, triangle $H$ = 4 cm, circle $H$ = 4.5
cm, and hexagon $H$=5 cm. The depth is measured from the initial
position of sand surface. }
\end{figure}

It can be seen from the Figs. 2 and 3 that the dependence of the
positions with the opening $h$ are very different for the two
holes. In the case of the first hole, $x_2$ values tend to remain
constant; the length of the hole $x_3 - x_1$, for the same depth
value $H$, increases for smaller values of the gate openings $h$
(Fig. 2); and the same applies to the depth $S$ (Fig. 4). On the
other hand, the second hole moves downstream when $h$ is smaller,
remaining practically constant the length of the hole $x'_3- x'_1
$ (Fig. 3). Moreover, the depth $S'$ also has a different
behavior. For a same depth value $H$, lower gate heights $h$
generate deeper holes until a threshold value, from which lower
value of $h$ generate holes less deeper (Fig. 4).  
As will be discussed in section 4, this phenomenon is caused by the influence of the first hole on the second one.

\subsection{Nondimensional numbers}

In order to gain clarity concerning the relation of the holes
positions with $h$ and $H$, we attempted to construct non-dimensional
numbers which could capture  this dependence in a simpler manner. We
considered different ways to define quantities having dimensions of
length, in order to non-dimensionalize the distances. The best
results were achieved with the quantities $\beta = H h^{-1} D^{-1}$
and $\gamma = h H^{-{{1} \over {2}}} D^{-{{3} \over {2}}}$, where $D$
is the width of the channel. In principle, the quantity $D$ is
unimportant to determine the flow, but it was included to get the
correct dimensions. In Figs. 5 and 6  the dimensionless
positions of the two holes , for different values of $\beta$ are
shown. It can be
seen that the non-dimensional positions fall fairly well on curves of
smooth variation.

\begin{figure}
\begin{center}
\includegraphics[width=1\textwidth]{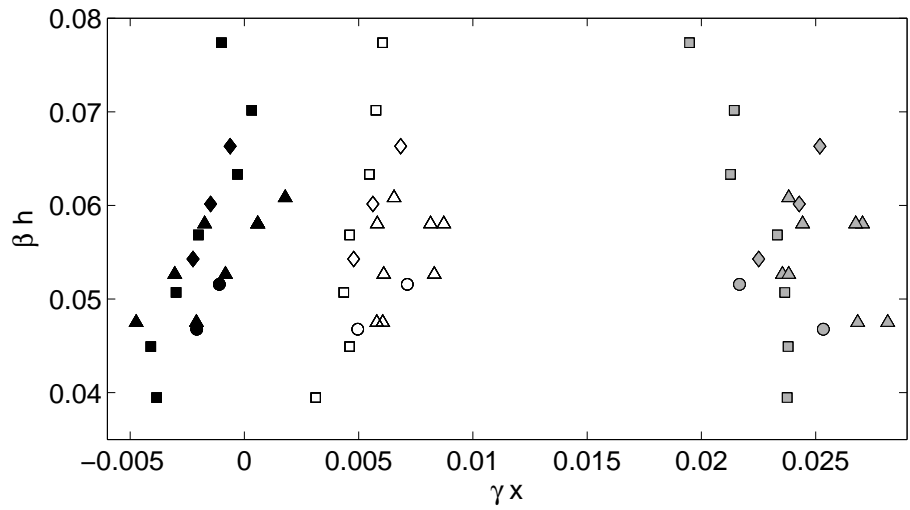}
\end{center}
\caption{Nondimensional positions ($\gamma x_1 $, $\gamma x_2 $,
$\gamma x_3 $) for the beginning (black), maximum depth (white)
and end (gray) of the first hole, for different values of the
nondimensionalized opening $\beta h$. The symbols shape are equal to
figure 2. }
\end{figure}

\begin{figure}
\begin{center}
\includegraphics[width=1\textwidth]{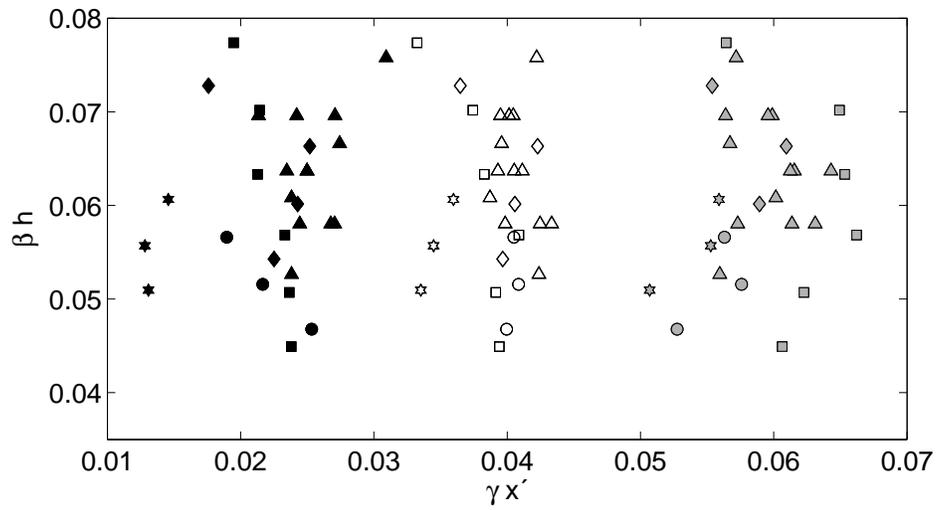}
\end{center}
\caption{Nondimensional positions ($\gamma x'_1 $, $\gamma x'_2 $,
$\gamma x'_3 $) for the beginning (black), maximum depth (white)
and end (gray) of the second hole, for different values of the
nondimensionalized opening $\beta h$. The symbols shape are equal to
figure 3. }
\end{figure}

\subsection{Flow structure}
The first hole has similar characteristics to the scour profile that
is usually observed downstream of submerged gates.
 This scouring is mainly driven by the jet that emerges from the
gate, which  is attached to the sediment bed.
  It is currently accepted that  erosion takes place when the shear
stress of the jet flow exerted on the sediment surface exceed a
   critical value which depends on the characteristic of the
particles. In wall jets, it is typically observed that the erosion
decreases downstream of the gate  with the
    distance due to the reduction of the jet velocity  (Hogg {\it et al.},1997).
    Thus it is intriguing that the second hole develops at positions
where the shear of the jet should not be sufficient for
     bed mobilization. We hypothesized that the process that cause the
second hole originates in the loss of the jet stability
    and the subsequent turbulence development. The
stability of a jet attached to a wall
      has been studied in many works ( Chun and Schwarz (1967), Bajura and Szewczyk (1970), Levin {\it et al.} (2005)) . These
studies showed that when the Reynolds number is larger than a
      critical value, the jet becomes unstable and the flow undergoes
a transition to turbulence.

In order to test this hypothesis, we implemented PIV measurements
and visualization by ink injection. In Fig. 7 the flow
field for $h=2$ cm, $H=4$ cm is presented, where the presence of
the jet is clearly seen. In Figs. 8 and 9 are shown the images
corresponding to experiments including fluid colored with ink.
From these figures, it is clearly seen that the region where there
is strong mixing produced by the turbulence, coincides with the
position of the second hole. The jet instability and the formation
of vortices can be observed in Fig. 10, which show the path
lines of the particles. These images support the view that it is
the turbulence developed after the destabilization of the jet that
originates the second hole. When the jet is separated from the sand
bed, vortices are formed between them. These vortices are shed
quasi-periodically and travel downstream. The dynamics is similar to
the one oberved in flows behind obstacles (Cabeza {\it et al.}, 2009).

Turbulent flow is not required to
develop scouring, but turbulence  increases erosion greatly
(Zanke, 2003). This is the reason why the second hole is developed
at a position where the laminar flow could not produce bed
displacement.

\begin{figure}
\hspace*{-28mm}
\includegraphics[width=1.4\textwidth]{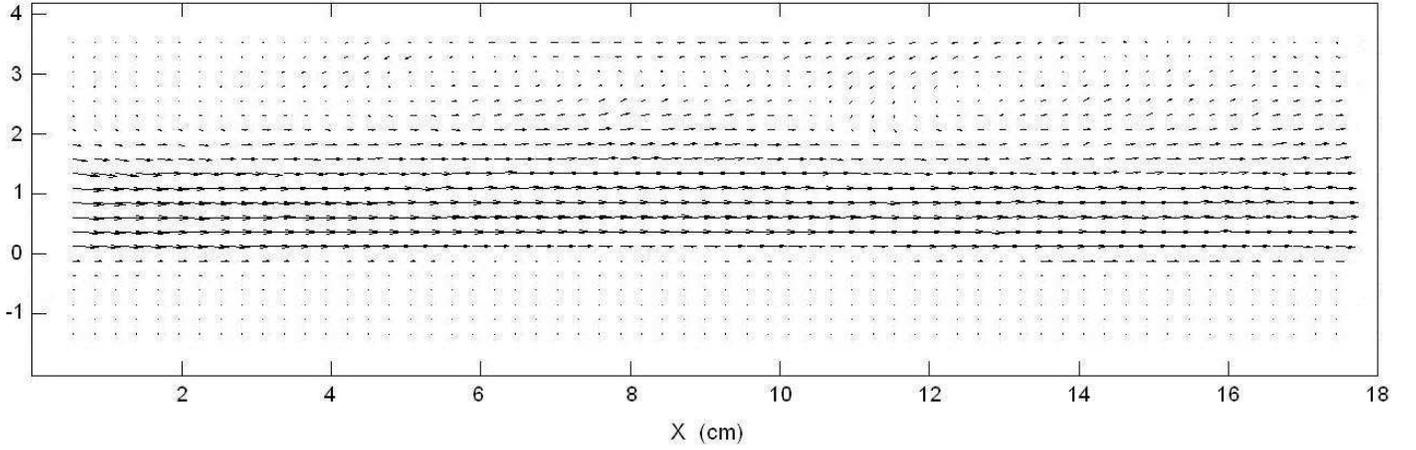}
\caption{Mean velocity field downstream the sluice gate obtained with
PIV technique, for $h=2$ cm and $H=4$ cm.}
\end{figure}

\begin{figure}
\vspace*{2mm}
\begin{center}
\includegraphics[width=1\textwidth]{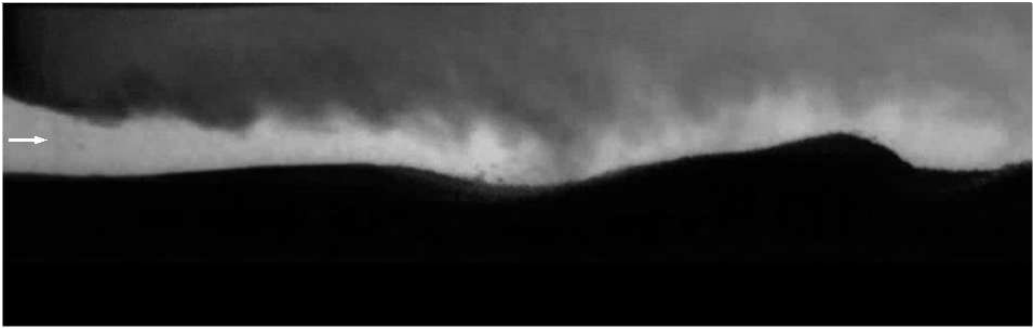}
\end{center}
\caption{Snapshot of the fluid flow, when the  upper layer was colored
with ink. It can be seen that at the position of the
second hole (at the center of the image) there is a strong mixing
between the two layers, due to the turbulence that arise
 from the destabilization of the jet. It also visible that  part of
the sediment is lifted due to turbulent bursts.}
\end{figure}

\begin{figure}
\vspace*{2mm}
\begin{center}
\includegraphics[width=1\textwidth]{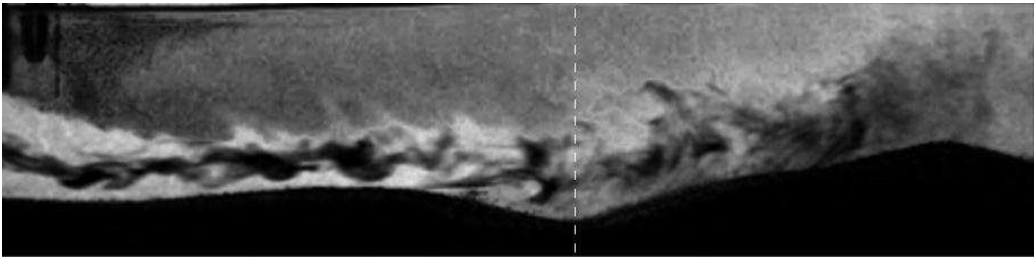}
\end{center}
\caption{Similar to figure 8, but now ink was injected also in the
jet. The dashed line is located on the second hole and is a guide to
the eye to remark that there is a change in the flow regime at this
position.}
\end{figure}

\begin{figure}
\vspace*{2mm}
\begin{center}
\includegraphics[width=1\textwidth]{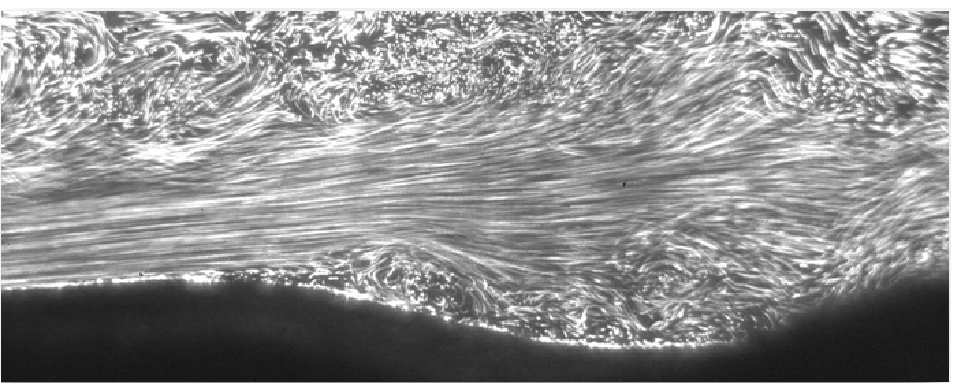}
\end{center}
\caption{Path lines of the particles illuminated by laser.}
\end{figure}

\section{Discusion}

The results about the flow structure show that the second hole
formation is related with the turbulence that develops after the
jet destabilization. When turbulence reaches the bed, strong
scouring takes place. We observed that this effect depends on the
evolution stage of the first hole. When the first hole is not
still well developed, the scouring at the position where the
second hole will appear is very weak. When the first hole begins
to form, the turbulent fluctuations at the bed surface become
significant and the depth of the second hole starts to grow. This
is probably due to the fact that the presence of the dune that
limits the first hole enhances the flow separation and the
instability of the jet. It is known that the wall exerts an
stabilizing effect on the jet flow. Thus, the formation of the
second hole, is favored with the existence of the first one. This
is true if the rate growing of the first hole is not too large. If
the first hole  forms too fast, the dune that forms downstream the
first hole alters the flow in a manner that the second hole is
inhibited. This situation is reached when the value of $h$ is
small. This tendency is reflected in Fig. 4. While the
larger values of the first hole depths $S$ are obtained with the
lowest values of $h$, the larger values of $S'$ correspond to
intermediate values of $h$. We notice that the formation of the
first hole is produced by the shear stress of the jet flow and
thus is not related to flow destabilization.

\section{Conclusions}

In this work we have studied the erosion produced by a turbulent jet
downstream of a sluice gate on noncohesive particles sediment bed.
We found new regimes in which two types of holes, clearly
distinguishable, are observed to develop. As the experimental results
shown, the
two holes are caused by distinct mechanisms.

The formation of the first hole is caused by the shear stress of the
jet flow and corresponds to the one usually observed
in eroded beds downstream gates. Then, the formation of the first hole
can be understood considering a laminar or a weak turbulence flow.

On the other hand, the second hole is
driven by turbulent fluctuations that stem from jet flow destabilization.
Since turbulence increases scouring greatly, the second hole is
developed at  locations where the laminar or weakly turbulent flow
could not produce bed displacement. However, in spite of the fact that
the holes originate from different mechanisms, the second hole
development depends on the first hole profile. The rate formation of
the second hole is maximized for a medium value of the first
 hole depth $S$. If $S$ is too large, the second hole is inhibited.
This implies that the maximum depth of the second hole takes place for
an intermediate value of the $h$ to $H$ ratio. In these conditions,
the depth of the second hole can be larger than those of the first
hole. As a consequence, the second hole is at least so important as is
the first one. Thus its study is of interest and deserves further
investigation.
We found nondimensional numbers for which the experimental data
concerning the positions of the holes and the ratio $h/H$ result ordered
in a simple way.

\section*{Acknowledgements}
\label{thanks}

This research was supported by CSIC and PEDECIBA, Uruguay.

\section*{References}

Bajura, R., Szewczyk, A. (1970). Experimental investigation of a laminar two-dimensional plane wall jet. {\it Phys. Fluids}. {\bf 13}, 1653$-$1664.

 \ \

\noindent Balachandar, R., Kells, J., Thiessen, R. (2000). The effect of tailwater depth on the dynamics of local scour. {\it Can. J. Civ. Eng.}. 27 {\bf 1}, 138$-$150.

\
 
\noindent Bey, A., Faruque, M., Baladanchar, R. (2007). Two-dimensional scour hole problem: role of fluid structures. {\it J. Hydraul. Eng.}. 133 {\bf 4}, 414$-$430.

\ \

\noindent Cecilia Cabeza, Juan Varela, Italo Bove, Daniel Freire,
Arturo C. Mart\'{\i},
     L. G. Sarasua, Gabriel Usera, Raul Montagne, and Moacyr Araujo (2009).
Two-layer stratified flows over pronounced obstacles at
low-to-intermediate Froude numbers. 
  {\it   Phys. Fluids } {\bf 21}, 044102 . 

\ \

\noindent Chatterjee, S., Ghosh, S., Chatterje, M. (1994). Local scour due to submerged horizontal jet. {\it J. Hydraul. Eng.}. 120 {\bf 8}, 973$-$992.

\ \ 

\noindent Chun, D., Schwarz, W. (1967). Stability of the plane incompressible viscous wall jet subjected to small disturbances. {\it Phys. Fluids}. {\bf 10}, 911$-$915.

\ \

\noindent D.F. Hill and B.D. Younkin (2009). 
A simple estimation of the growth rate and equilibrium size of bedforms
created by a turbulent wall jet.
{\it J. Hydraul. Res}. {\bf 47},  (2009), 619$-$625.
 
\ \

\noindent Hogg, A., Huppert, H., Dade, W. (1997). Erosion by planar turbulent jets. {\it J. Fluid. Mech.}. {\bf 338}, 317$-$340, and references therein.
\ \

\noindent Levin, O., Chernoray, V., Lofdahl, L.,Henningson, D. (2005). A study of the Blasius wall jet.  {\it J. Fluid Mech.}. {\bf 539}, 313$-$347.

\ \
 
\noindent Rajaratnam, N. (1981). Erosion by plane turbulent jets. {\it J. Hydraul. Res}. 19 {\bf 4}, 339$-$358.

\ \ 

\noindent Zanke, U. (2003). The influence of turbulence on the initiation of sediment motion. {\it International Journal of Sediment Research}. 1 {\bf 18}, 17$-$31.

\end{document}